\documentclass[conference]{IEEEtran}
\IEEEoverridecommandlockouts
\usepackage{cite}
\usepackage{amsmath,amssymb,amsfonts}
\usepackage{graphicx}
\usepackage{textcomp}
\usepackage{xcolor}

\usepackage{amsmath, amssymb, amsthm,algorithm}
\usepackage[latin1]{inputenc}
\usepackage{xcolor}
\usepackage{latexsym}
\usepackage{graphics,epsfig}
\usepackage{amsfonts}
\usepackage{booktabs}
\usepackage{color}
\usepackage{multirow}
\usepackage[justification=centering]{caption}
\usepackage{graphicx}
\usepackage{adjustbox}
\usepackage{kantlipsum}
\usepackage[caption=false,font=footnotesize]{subfig}
\usepackage{cases}
\usepackage{url}
\usepackage{tabu}
\usepackage{makecell}
\usepackage{soul}

\usepackage{array}
\usepackage{algpseudocode}
\usepackage[justification=centering]{caption}

\usepackage[caption=false,font=footnotesize]{subfig}
\usepackage[T1]{fontenc}
\usepackage{mwe}
\usepackage{subfig}
\usepackage{mathtools,lipsum,cuted}

\usepackage[english]{babel}
\usepackage{verbatim}

\usepackage[english]{babel}

\def\BibTeX{{\rm B\kern-.05em{\sc i\kern-.025em b}\kern-.08em
    T\kern-.1667em\lower.7ex\hbox{E}\kern-.125emX}}
\begin{document}

\title{Performance Analysis of MDPC and RS codes in Two-channel THz Communication Systems}
\author{\IEEEauthorblockN{Cao Vien Phung$^{1}$, Christoph Herold$^{2}$, David Humphreys$^{3}$, Thomas K\"urner$^{4}$ and Admela Jukan$^{5}$}
\IEEEauthorblockA{$^{1,5}$Institut f\"ur Datentechnik und Kommunikationsnetze, Technische Universit\"at Braunschweig, Germany\\
$^{2,4}$Institut f\"ur Nachrichtentechnik, Technische Universit\"at Braunschweig, Germany\\
$^{3}$National Physical Laboratory, Teddington TW 11 0LW, United Kingdom \\
Email: \{$^{1}$c.phung, $^{5}$a.jukan\}@tu-bs.de  and \{$^{2}$herold, $^{4}$kuerner\}@ifn.ing.tu-bs.de
 and $^{3}$david.humphreys@npl.co.uk}}
\maketitle

\begin{abstract}
We analyze whether a multidimensional parity check (MDPC) or a Reed-Solomon (RS) code in combination with an auxiliary channel can improve the throughput and extend the THz transmission distance. While channel quality is addressed by various coding approaches, and  an effective THz system configuration is enabled by other approaches with additional channels, their combination is new with the potential for significant improvements in quality of the data transmission. Our specific solution is designed to correct data bits at the physical layer by using a low complexity erasure code (MDPC or RS), whereby original and parity data are transferred over two separate and parallel THz channels, including one main channel and one additional channel. The  results are theoretically analyzed to see that our new solution can improve throughput, support higher modulation levels and transfer data over the longer distances with THz communications.
\end{abstract}

\section{Introduction} \label{intro}
The terahertz (THz) frequency band ($0.3$-$10$ THz) is envisioned to meet the needs of ever increasing multi-gigabit mobile data in future wireless networks \cite{6248357}. Communications in the THz frequency range are however very susceptible to atmospheric effects due to gaseous absorption and water vapor. Depending on the transmission frequency chosen, these effects can have a major impact on the feasibility of high-data rate communications \cite{itup676, itup838, itup840}. A careful selection of the channelization is therefore required to mitigate these impairments. Transmitting data with a high modulation level over a long transmission distance in this frequency range can lead to a deteriorated Signal-to-Noise ration (SNR), i.e., a high BER (Bit Error Ratio), SER (Symbol Error Ratio) or block error ratio of transmitted data. A large bandwidth and thus short symbol durations present a challenge for traditional Forward Error Correction (FEC) codes. The adaption and development of coding schemes to provide the desired robustness for fast and reliable high data rate transmission are thus subject of ongoing research \cite{2018_WEHN.SAHIN_NextGenerationChannelCoding}. The IEEE 802.15.3d standard \cite{8066476} gives guidance on the implementation of high data rate transmissions in the low THz range of 252 GHz to 325 GHz. The research community is trying to address the challenges related to THz communication for both system configuration issues and new transceiver design engineering.

The usage of auxiliary channels in combination with coding provide a quite an interesting solution idea to achieve robustness for the data transmission towards better quality of transmission. A Wi-Fi auxiliary channel, combined with THz transmission channel, was used to estimate distance of the THz receiver and measure air humidity \cite{7444891}. Ultra-Massive MIMO Systems with auxiliary channels, deployed at THz band, can solve the transmission distance issue and improve the capacity of system \cite{faisal2020ultramassive}. On the other hand, to protect original data against losses, erasure codes with redundant data are proposed to improve the quality of the data transmission. Today, a few FEC codes, such as Reed-Solomon (RS), Low Density Parity Check (LDPC) or Polar code \cite{sarieddeen2020overview,wehn_norbert_2018_1346686, wehn_norbert_2019_3360520} are still under discussion. Whereas the usage of auxiliary channels or coding alone is not new, the combination of these techniques in THz communications is novel and promising, as it can increase the system capacity from auxiliary channels and at the same time significantly improve transmission quality. 

This paper analyzes the performance in a system that combines an auxiliary THz channel and a multidimensional parity check (MDPC) code \cite{Shea2005MultidimensionalC,8867037} or RS code \cite{Reed1999} with goal of improving the THz transmission distance and transmission quality. Depending on the achievable code rate, the transmission of redundancy bits over independent, auxiliary channels allows for the usage of more robust transmission parameters due to a lower required data rate. This concept offers additional layer of coding to further improve the system's BER performance. MDPC and RS codes are chosen because they are systematic codes with reasonable simplicity and low complexity. The overall solution presents a good choice as the source data keeps unchanged during the encoding process at the sender, which additionally reduces coding and decoding overhead. Furthermore, the feature of systematic codes is easy to execute for multi-channel systems, e.g., the data can be configured and sent over the main channel without waiting for the coding process by the sender. The additional parity bits can be carried by the auxiliary channel to support the decoding process by the receiver, so an increase of the modulation rate is not required. The data of each coding block is sent over two parallel channels, whereby original data of each block is sent over the main THz channel, while parity bits created during the encoding process are transferred over the auxiliary THz channel.The results demonstrate higher fault tolerance and overall throughput with a higher modulation level and a longer transmission distance.
\begin{figure*}[!ht]
\centering
\includegraphics[width=1.9\columnwidth]{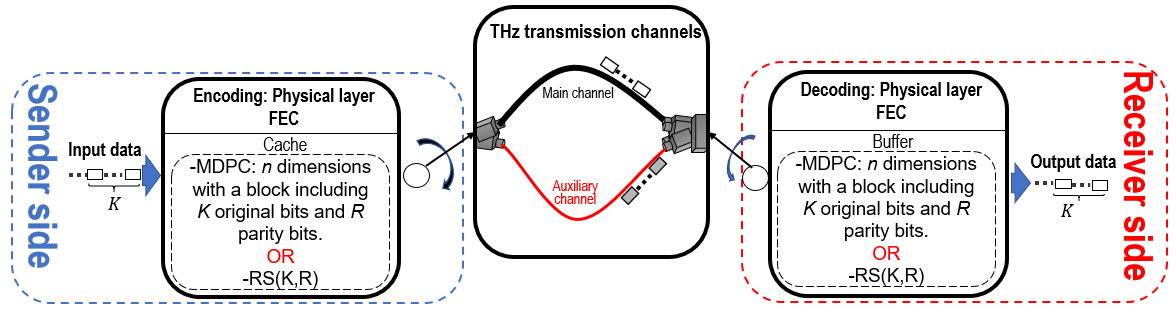}
  \vspace{-0.1cm}
  \caption{Two-channel THz system with MDPC or RS codes.}
  \vspace{-0.6cm}
  \label{scenario}
\end{figure*} 

\section{System design} \label{sysde}
Fig. \ref{scenario} shows the basic reference architecture of two-channel THz system. Independently whether RS or MDPC coding is used over both the main and auxiliary channel. Source data is first encoded to generate coding redundancy, and then distributed over two channels to be simultaneously sent to the receiver. Both main and auxiliary channel use THz technique, whereby the auxiliary THz channel is utilized for transferring the coding redundancy, while the main THz channel is used for sending the original data. Since the amount of original data is more than or equal to the amount of redundant data, the transmission rate of the auxiliary channel should be lower than or equal to that of main channel. To restrict the issue of co-channel interference, auxiliary and main channel should be spatially divided or used at different frequencies. The lower frequency should be used for the auxiliary channel, while the higher one should be configured for the main channel; this will be discussed later in section \ref{numre}. To achieve a high bit rate, we can use high-order modulations because the quality of the data transmission can be satisfied by the redundancy sent over the auxiliary channel.


In Fig. \ref{scenario}, any input data represented by a long bit stream is split into substreams of $K$ bits each. The coding cache is created to temporarily  store any $K$ bits arrived at the source for coding process. A block is defined to be a set of bits coded and decoded with each other, e.g., each block has $K$ original bits and redundant data $R$.  We consider a MDPC code, labeled as MDPC($n$D/$m$L), with $n$ dimensions (D) and the same length (L) of $m$ data bits for each dimension or  RS code.

\subsubsection{MDPC code} With MDPC($n$D/$m$L) code, based on \cite{8867037}, each input substream is equal to $K=m^n$, whereby  $m$ and $n$ are chosen for $K$ under Eq. \eqref{parameters-mn} to optimize the two-channel THz system with fault tolerance as discussed in section \ref{ana}. The encoding process of the encoder for the parity bits can be referred to \cite{8867037}: A MDPC code with $n=2$ dimensions is a special case of multidimensional codes, whereby data bits are put into a two-dimensional matrix including columns and rows, whereby column and row parity bits are placed in the last row and column, respectively; each column parity bit in the last row secures data bits in its column and each row parity bit in the last column secures data bits in its row; the parity check bit on check bits for securing column and row parity bits is placed at the bottom right corner of the matrix. For $n \geq 3$, MDPC code is a construction of combining the basic three-dimensional parity check cubes into a multidimensional hypercube, whereby a three-dimensional parity check cube is constructed by putting the two-dimensional parity matrices layered into a third dimension and the parity bits across the layers of data bits belong to the last layer. However, $n=2$ is a good choice for this system in terms of code rate, which will be discussed in section \ref{numre}. After the coding processing, the encoder generates $R$ parity bits, i.e., $K+R=(m+1)^n$ bits in total \cite{8867037}, so $R=(m+1)^n-m^n$ parity bits. We note that the $K$ original bits stay unchanged during the coding process.

\subsubsection{RS code}  Based on \cite{Reed1999}, we apply the following: The redundant data $CK(X)$ is generated from the original data $M(X)$ with the size of $K$ bits by the modulo-$g(X)$ function, i.e., $CK(X)=X^{\frac{R}{s}}\cdot M(X) \; mod \; g(X)$, whereby $X^{\frac{R}{s}}$ is the displacement shift, $R$ is the size of parity-check information bits and $g(X)$ represents the generator; and then the code word $C(X)$, including the redundant data $CK(X)$ appended systematically onto the original data, i.e., $C(X)=X^{\frac{R}{s}}\cdot M(X)+CK(X)$, will be sent over transmission channels; as a result, each block of $K$ original information bits will be increased to $R+K$ bits after the coding process, whereby $R+K$ bits are divided into symbols of $s$ bits each, all operations are performed via finite field $\mathbb{F}_{2^s}$ and original bits are unchanged during the coding process. The parameters of $K$, $R$ and $s$ are chosen under Eq. \eqref{parameters-RKforRS} in section \ref{ana} to optimize the two-channel THz system with fault tolerance, whereby the maximum codeword length is $\frac{K+R}{s}=2^s-1$. Assume the value of $R$ is fixed, if  $\frac{K+R}{s}<2^s-1$, the encoder can add $Z$ zero padding symbols into $\frac{K}{s}$ original symbols so that the codeword length can achieve $\frac{K+R}{s} + Z=2^s-1$, and then the coding process can be performed on this new one for generating $\frac{R}{s}$ parity-check symbols. We note that the zero padding symbols will not be sent over transmission channels, but the decoder at the receiver side have to add these ones for the decoding process.

Finally, any $K$ original bits and the $R$ parity bits are sent in parallel over the main and auxiliary channel, respectively. In our evaluation, we consider these channels to be independent transmission channels without co-channel interference.  

\subsubsection{Decoding} For decoding, we need to collect sufficient data and parity bits from the same block on both channels. For MDPC($n$D/$m$L) code, the decoder uses an iterative decoding algorithm proposed by \cite{8867037}, whereby the order of bits in transmission is also known by the decoder at the receiver for arranging them into a multidimensional matrix for the decoding process. The  RS code \cite{Reed1999} decoding process follows a five-stages: The first stage is to calculate the syndrome components from the received word; the second stage is for calculating the error-locator word from the syndrome components; we calculate the error locations from the error-locator numbers which are from the error-locator word for the third stage; the fourth stage is the calculation of the error values from the syndrome components and the error-locator numbers; and the final stage in the process is the calculation of the decoded code word from the received word, the error locations, and the error values. The function of MDPC($n$D/$m$L) or RS code can detect and correct error bits. The receiving buffer used to temporally store all arriving data for the decoding process is always a major challenge in high speed systems. 

\section{Analysis} \label{ana}
This section shows the analysis for two-channel THz transmission system with MDPC($n$D/$m$L) or RS code.
\subsection{THz system optimized for fault tolerance}
\subsubsection{Analysis of optimized value $K$ for MDPC($n$D/$m$L) code}
In general, according to \cite{939851}, the minimum Hamming distance of MDPC($n$D/$m$L) code is $\Delta_{min }=2^n$. Hence, the number of error bits that can be corrected by MDPC($n$D/$m$L) code is:
\begin{equation}\label{t_k}
t_{MDPC}=\frac{\Delta_{min}-2}{2}=\frac{2^n - 2}{2 }=2^{n-1} - 1
 \end{equation}
 
 As discussed in \cite{8867037}, with any MDPC($n$D/$m$L) code, the number of data bits at the input encoding process of the sender is $K=m^n$ sent over the main THz channel with bit error probability $p_e^M$ and $R=(m+1)^n - m^n$ parity bits generated by the encoder are transferred over the auxiliary THz channel with bit error probability $p_e^A$. We need to choose $m$ and $n$ ($m,n \in \mathbb{N}$) so that the total number of error bits occurred when transmitting $K+R$ bits over two THz channels is lower than or equal to the ability of correcting bit errors by MDPC($n$D/$m$L) code to optimize the system with fault tolerance, i.e., 
  \begin{equation}\label{parameters-mn}
m^n \cdot p_e^M \leq t_{MDPC},
 \end{equation}
 where $m \geq 0$ and the transmission distance of auxiliary channel is assumed with the setting so that $p_e^A=0$.
 \subsubsection{Analysis of optimized value $K$ for RS code}
The minimum Hamming distance of RS code in symbols \cite{Reed1999} is generally given as: $\Delta_{min}=\frac{R}{s} + 1$. The number of error symbols corrected by RS code can be given as:
\begin{equation}\label{t_k_RS}
t_{RS}=\frac{\Delta_{min}-1}{2}=\frac{R}{2s},
 \end{equation}
 where an error symbol is defined to be contained at least one damaged bit, i.e., an error symbol has the number of error bits in $[1,s]$. The expected SER on the main channel is:
\begin{equation}\label{p_s}
P^M_s=1-(1-p^M_e)^{s}
 \end{equation}
where $p_e^M$ is the expected BER on the main channel and the transmission distance of auxiliary channel is set so that $p_e^A=0$, i.e., The expected SER $P^A_s$ on the auxiliary channel is equal to $0$. As $K$ native symbols sent over the main channel with symbol error probability $P_s^M$, and $R$ parity symbols sent over the auxiliary channel with  symbol error probability $P_s^A=0$, we choose $K$, $R$ and $s$ so that they are sent over two THz channels in total, then the total number of error symbols received by the receiver should be lower than or equal to the ability of correcting symbol errors by RS coding mechanism, with the aim of optimizing the fault tolerant system, and the maximum codeword length is $\frac{K+R}{s}=2^s-1$, i.e.,
 \begin{equation}\label{parameters-RKforRS}
\text{      } \left\{\begin{matrix}
\frac{K}{s} \cdot  P_s^M    \leq t_{RS}\; \; ;\; \; K\geq 0 & \\ 
2^{s-1}\leq \frac{K}{s}+\frac{R}{s} \leq 2^s - 1&,
\end{matrix}\right.
 \end{equation}
 \subsubsection{Code rate, transmission overhead and information bit rate for  MDPC($n$D/$m$L) and RS code }
Based on Eq. \eqref{parameters-mn} for MDPC($n$D/$m$L) code and Eq. \eqref{parameters-RKforRS} for RS code, the code rate $R_F$ of MDPC($n$D/$m$L) or RS code can be expressed by:
  \begin{equation}\label{coderate}
R_F=\frac{K}{K+R}.
 \end{equation}
 
With Eq. \eqref{coderate}, the transmission overhead $\theta$ of two-channel THz system can be given as:
 \begin{equation}\label{overhead}
 \theta = 1-R_F.
 \end{equation}
 
\subsection{Block error probability of MDPC($n$D/$m$L) and RS code}
In this subsection, we analyze the block error probability of MDPC($n$D/$m$L) and RS code. We assume that any arbitrary block of $K$ original bits chosen is sent over the main channel, $R$ parity bits generated from $K$ original data are transferred over the auxiliary channel, where the bit error probability of all channels is more than or equal to $0$. We define an error block to be at least one bit of it damaged.



 \begin{figure*}[ht]
  \centering
  \hfil \hfil \hfil \hfil \hfil \hfil
  \subfloat[Bandwidth of $2.16$ GHz at center frequency $294.84$ GHz.]{\includegraphics[ width=6cm, height=5cm]{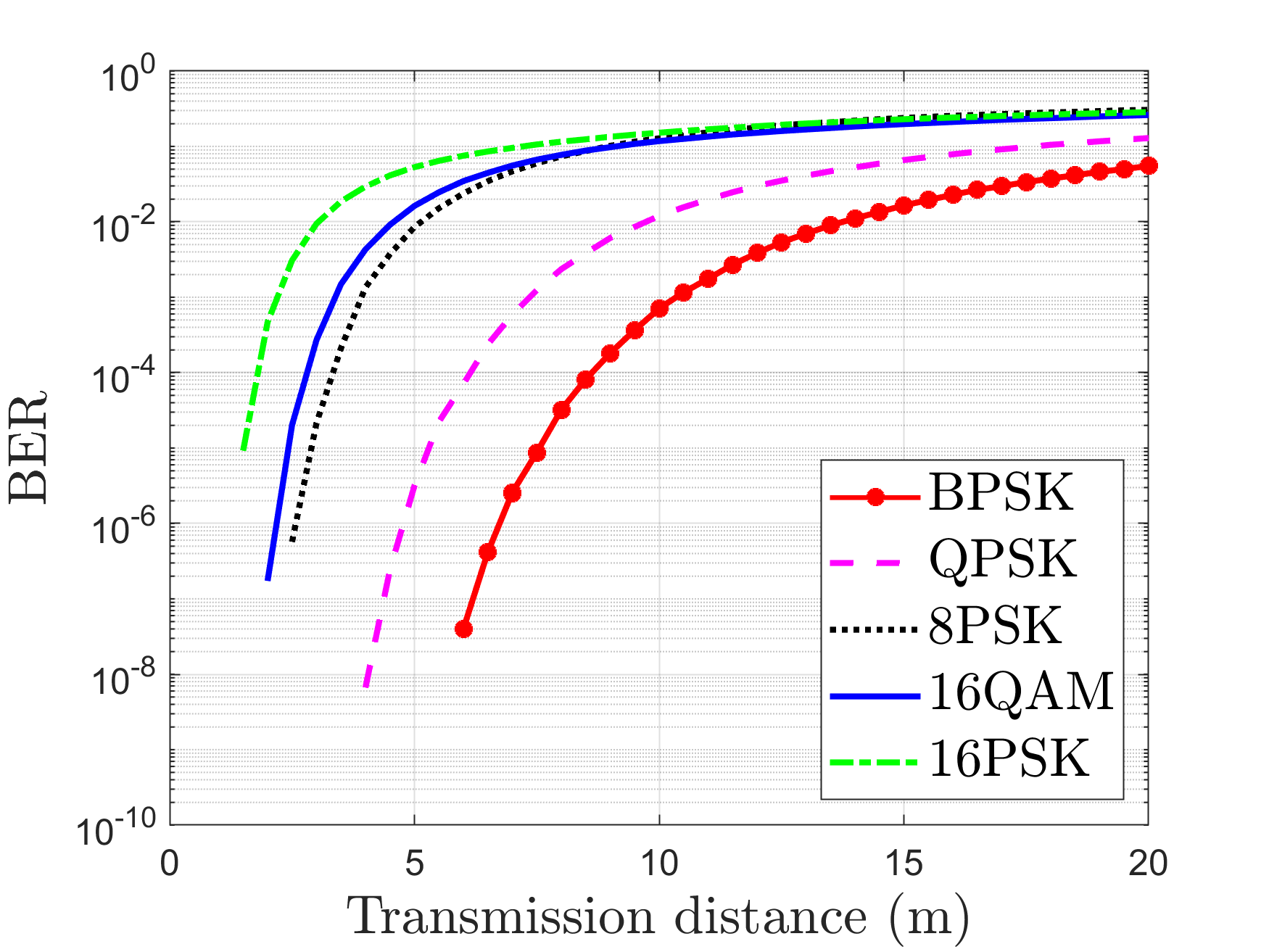}
  \label{294dot84}}
  \subfloat[Bandwidth of $8.64$ GHz at center frequency $300.24$ GHz.]{\includegraphics[ width=6cm, height=5cm]{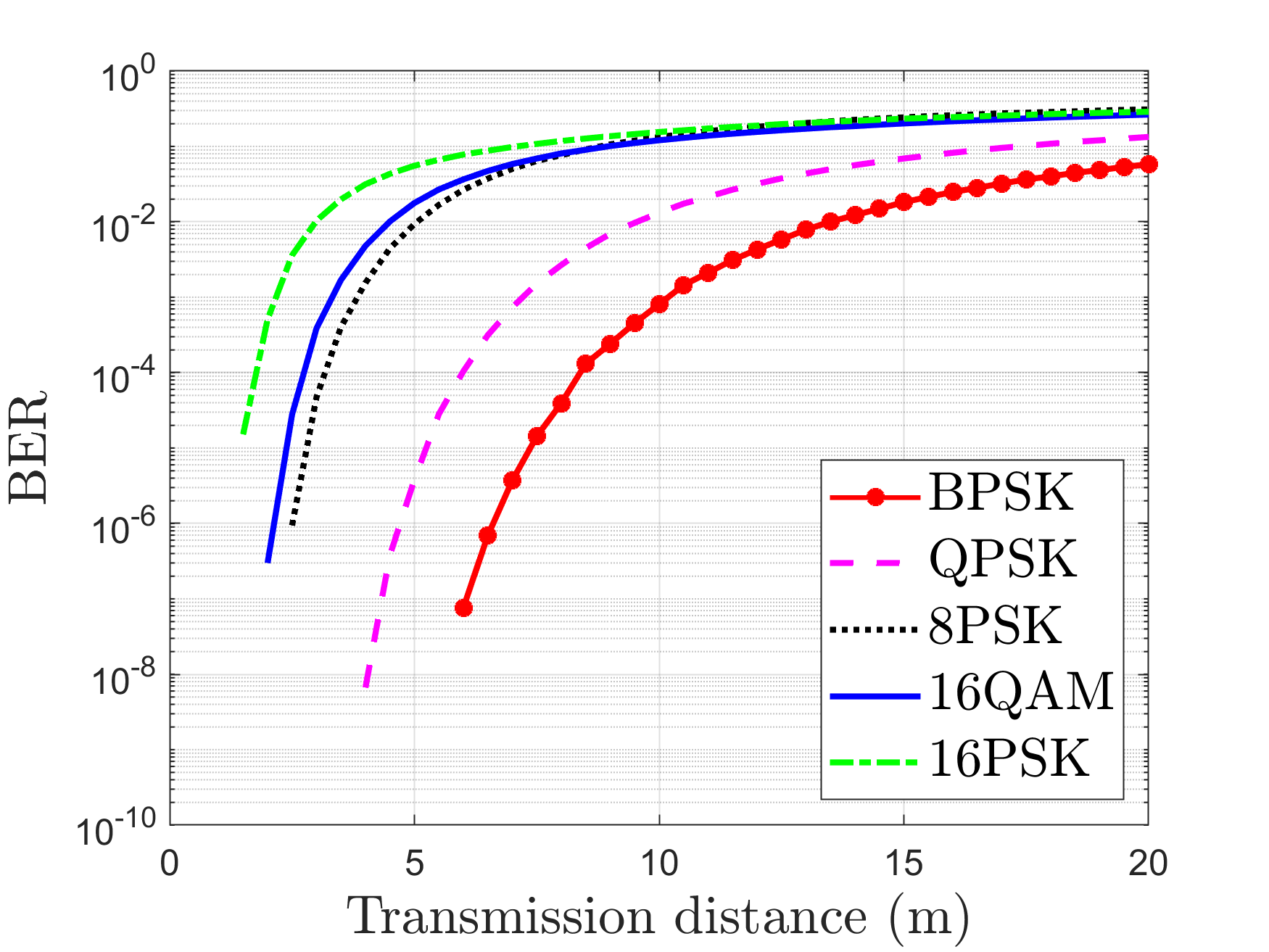}
  \label{BW8_64_300dot24}} 
  \subfloat[Bandwidth of $10.80$ GHz at center frequency $299.16$ GHz.]{\includegraphics[width=6cm, height=5cm]{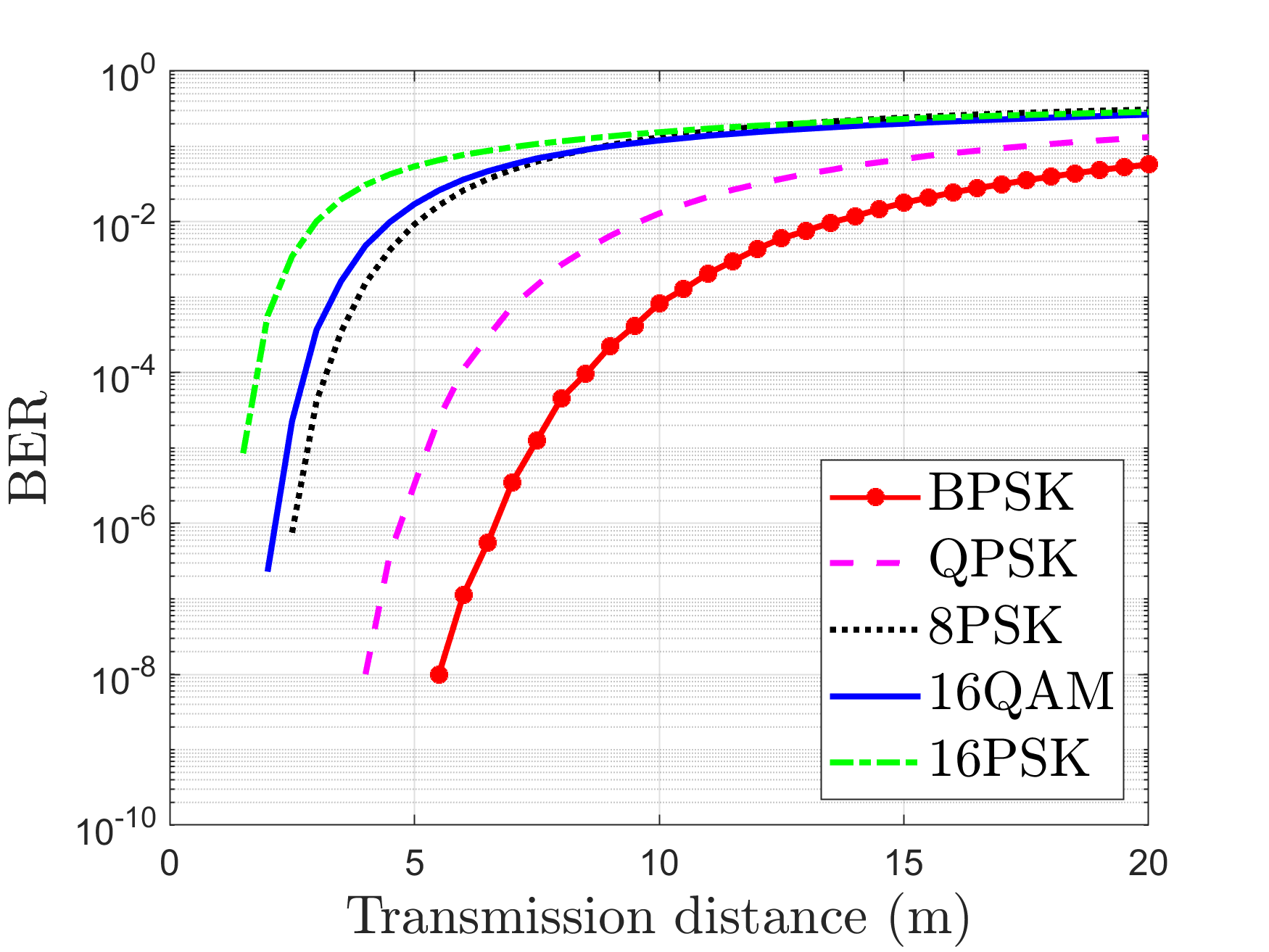}
  \label{BW10_80_299dot16}}
  \caption{BER vs. transmission distance without coding.}
  \label{mixfrequency}
  \vspace{-0.4cm}
  \end{figure*}
  
    \begin{figure}[ht]
  \centering
  \includegraphics[width=\columnwidth]{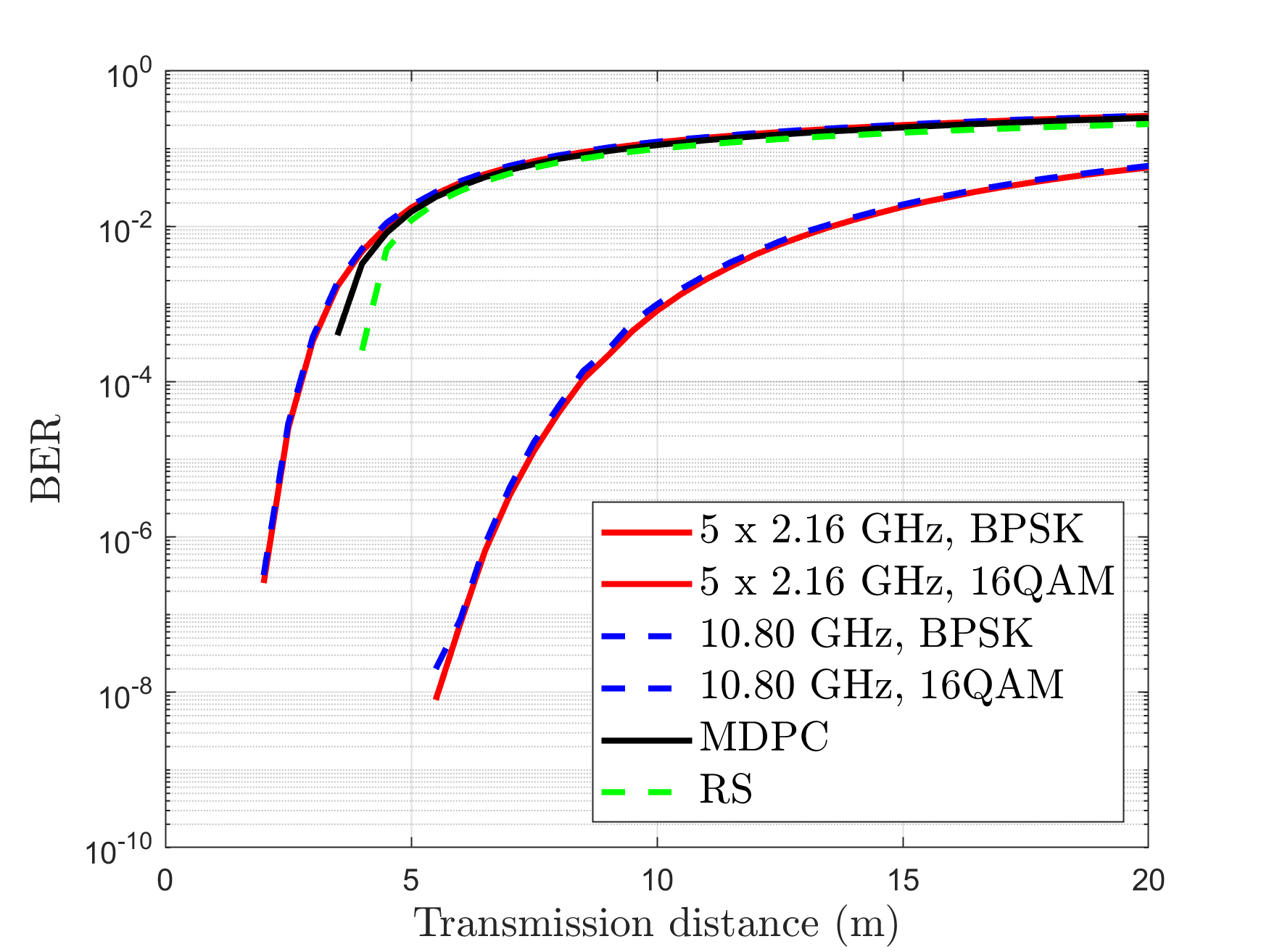}
  \caption{Bit error rate of all three systems with the same $10.80$ GHz total bandwidth and equal transmit power of $-8$.}
  \label{BER_comparison}
  \vspace{-0.4cm}
  \end{figure}
  
   \begin{figure}[ht]
  \centering
  \includegraphics[width=\columnwidth]{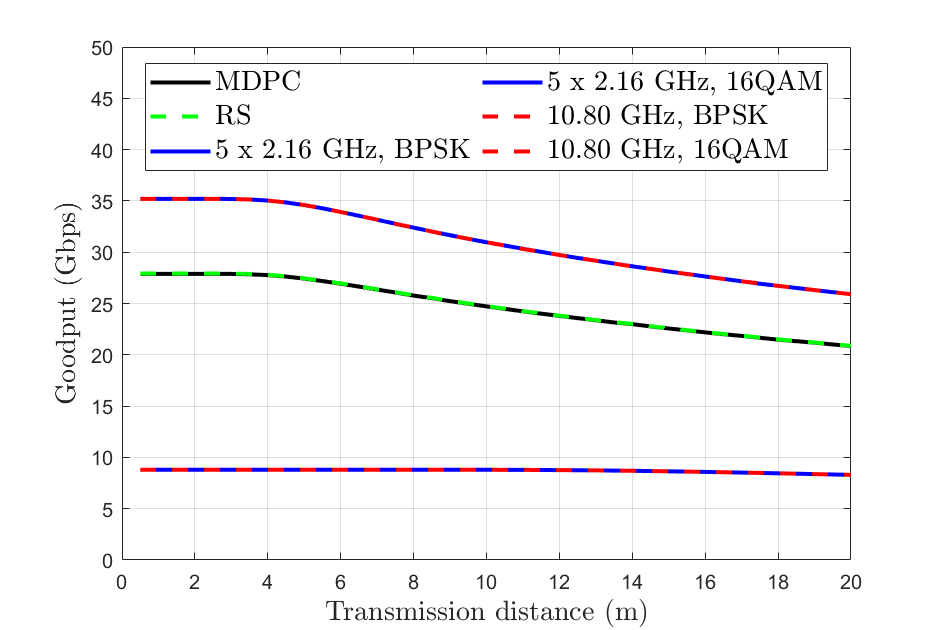}
  \caption{Throughput of successfully transmit information bits (goodput) of the three compared systems}
  \label{Goodput_comparison}
  \vspace{-0.5cm}
  \end{figure}
  
       \begin{figure*}[ht]
  \centering
  \hfil \hfil \hfil \hfil \hfil \hfil
  \subfloat[System with BER of auxiliary channel unequal to $0$ and code rate $R_F=0.93$.]{\includegraphics[ width=6cm, height=5cm]{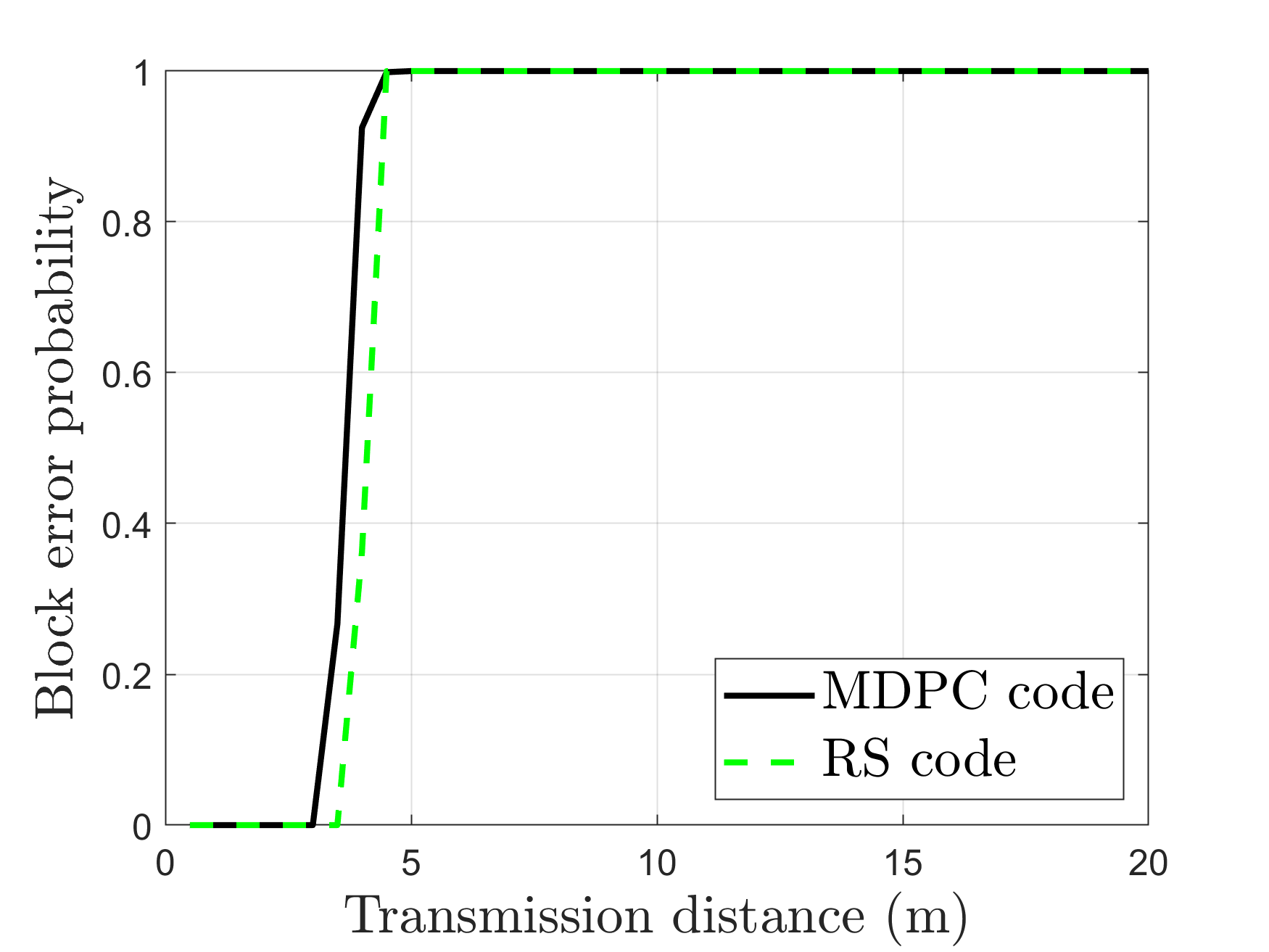}
  \label{error_block_for_coding}}
  \subfloat[System with BER of auxiliary channel equal to $0$.]{\includegraphics[ width=6cm, height=5cm]{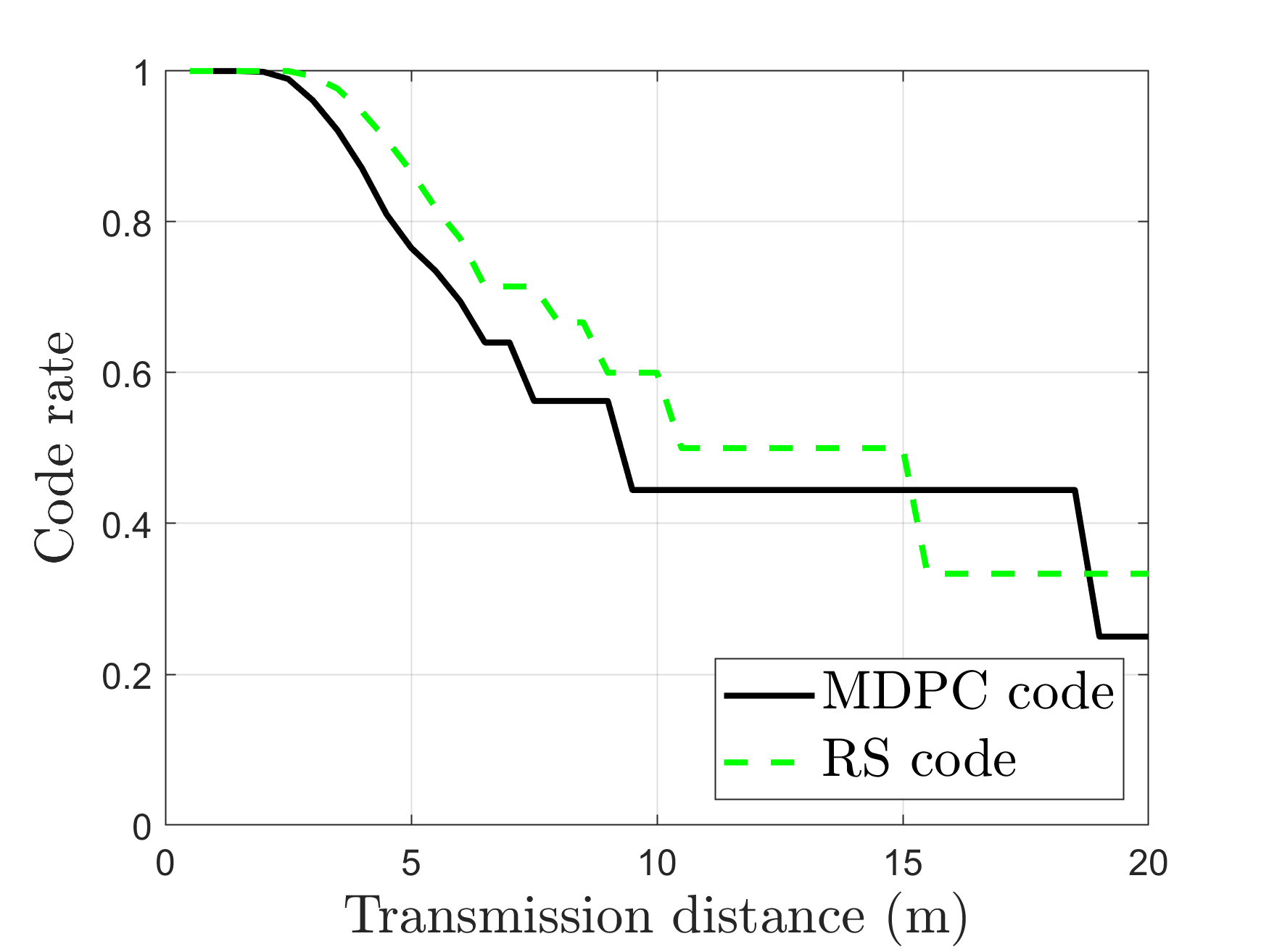}
  \label{coderate_eve}}
  \subfloat[System with BER of auxiliary channel equal to $0$.]{\includegraphics[ width=6cm, height=5cm]{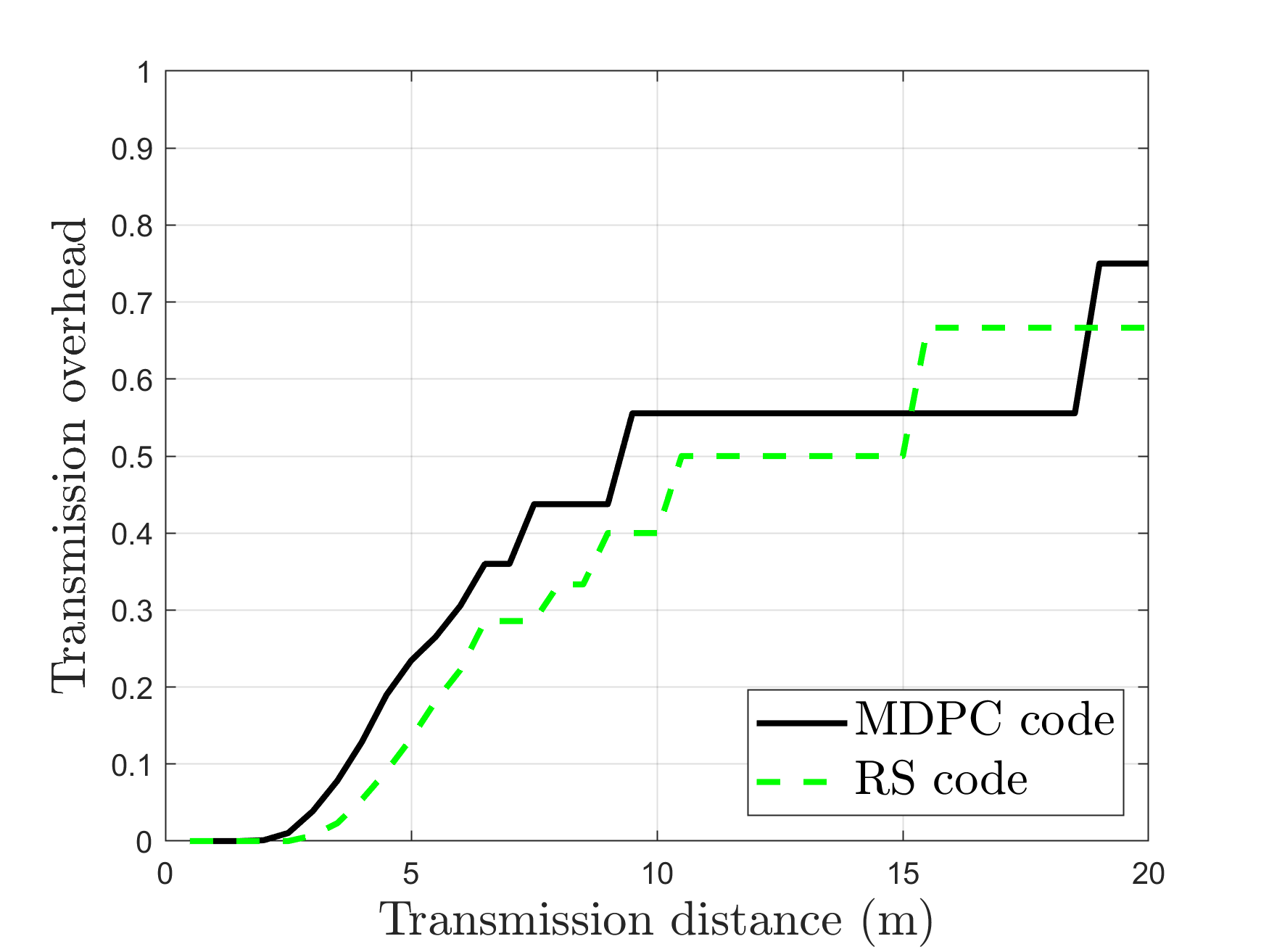}
  \label{trans_overhead}} 
  \caption{Block error probability, code rate and transmission overhead, for two-channel THz system with coding.} 
  \label{erorblock-coderate-transover}
 \vspace{-0.4cm}
  \end{figure*}

\subsubsection{Block error probability for MDPC($n$D/$m$L) code}
With each certain transmission distance of main channel, $d_{main}$, and auxiliary channel, $d_{aux}$, the corresponding BER is $p_e^M$ and $p_e^A$, respectively. Using Eq. \eqref{t_k}, the expected residual BER of MDPC($n$D/$m$L) code after the decoding process performed by the receiver can be given as:
 \begin{equation} \label{P_r}
 P_{re}^{MDPC} = \frac{ K \cdot p_e^M +  R\cdot p_e^A  -t_{MDPC}}{K+R},
 \end{equation}
where $P_{re}^{MDPC} \leq 0$ means MDPC($n$D/$m$L) code can correct all bit errors. Hence, if $P_{re}^{MDPC} < 0$, then we assign $P_{re}^{MDPC} = 0$. Hence, using Eq. \eqref{P_r}, at the receiver side, the block error probability of MDPC($n$D/$m$L) code after the decoding process can be given as:
\begin{equation}\label{P_b}
P_b^{MDPC} = 1 - (1-P_{re}^{MDPC})^K.
\end{equation}
\subsubsection{Block error probability for RS code}
With each certain transmission distance of main channel, $d_{main}$, and auxiliary channel, $d_{aux}$, the corresponding SER is $P_s^M$ and $P_s^A$, respectively.  Using Eq. \eqref{t_k_RS}, the expected residual SER of RS  code after the decoding process performed by the receiver is:
 \begin{equation} \label{P_r_RS}
 P_{rs}^{RS} = \frac{ \frac{K}{s} \cdot P_s^M +  \frac{R}{s}\cdot P_s^A  -t_{RS}}{\frac{K+R}{s}},
 \end{equation}
 where $P_{rs}^{RS} \leq 0$ means RS code can correct all symbol errors. Hence, if $P_{rs}^{RS} < 0$, then we assign $P_{rs}^{RS} = 0$. Hence, the expected residual BER for RS code is calculated as:
  \begin{equation} \label{P_re_RS}
 P_{re}^{RS} = 1-\sqrt[s]{1-P_{rs}^{RS}}
 \end{equation}
 
  Hence, using Eq. \eqref{P_r_RS}, at the receiver, the block error probability of RS code after the decoding process is:
 \begin{equation}\label{P_b_RS}
P_b^{RS} = 1 - (1-P_{re}^{RS})^{K}.
\end{equation}

\section{Numerical results} \label{numre}
The IEEE 802.15.3d standard foresees the frequency range of $252$ GHz to $325$ GHz for high data rate wireless communication and a channelization of wireless communication channels with eight different bandwidths between $2.16$ GHz and $69.12$ GHz \cite{8066476}. For the evaluation of the proposed system, a total of $10.80$\,GHz around the center frequency of 299.16 GHz and a transmission power of $-8$\,dBm are used. By employing a part of the available spectrum to form an auxiliary channel, additional protection of the transmitted data can be reached by transferring redundant bits over a separate channel with a more robust transmission. The selection of the bandwidth and modulation scheme of main and auxiliary channel should reflect the chosen code rate to facilitate the synchronization of both channels. The (240;224)-RS-code specified in the IEEE802.15.3d standard for example, can be constructed by splitting up the $10.80$\,GHz into an auxiliary channel with a BPSK (1 bit/symbol) modulation scheme and a 2.16 GHz bandwidth (at the center frequency $294.84$ GHz and a main channel using a 16 QAM (4 bits/symbol) and 8.64 GHz bandwidth (at a center frequency of $300.24$ GHz). The combination of different bandwidths and modulation schemes produces a 16:1 ratio between main and auxiliary channel, which is close to the 14:1 ratio of the coding scheme. The difference between the ratios could be bridged by zero padding the main channel's bit stream, for example. 

For the evaluation, this (240;224)-RS code and a comparable MDPC($n$D/$m$L) code with $n=2$ dimensions and $784$ information bits and $841$ coded bits have been used. Both have a code rate $R_F = 0.93$. The proposed system is contrasted to two different communication setups that are further referred to as reference systems: Five separate channels of 2.16 GHz bandwidth and the same modulation scheme (at the respective center frequencies $294.84$ GHz, $297.00 $ GHz, $299.16$ GHz, $301.32$ GHz and $303.48$ GHz) and a single channel of 10.80 GHz (at the center frequency of $299.16$\,GHz). The transmit power is spread evenly over the channels, meaning that each of the five 2.16 GHz bandwidth channel will have only one fifth of the -8 dBm transmit power of the 10.80 GHz band. The code rate $R_F$ of both systems is $R_F = 1$. The BERs for the comparison of the channels are based on simulations conducted using using the Simulator for Mobile Networks (SiMoNe) \cite{9391508} and considering free space path loss over distances between $0.5$ m and $20.0$ m in steps of $0.5$ m. Due to the short distances and the selection of suitable frequency ranges with to low atmospheric attenuation, atmospheric effects have not been simulated. A link budget has been created assuming a transmit power of $-8$ dBm, an antenna gain of $26.4$ dBi -- both at transmitter and receiver. A root-raise cosine pulse with a roll-off factor of $\alpha = 0.4$ is used for the pulsed transmission. Noise is modeled as thermal noise and added to the transmission signal. The simulated thermal noise has a noise temperature of $290$ K. The receiver's noise figure is chosen to be $10$ dB. To show the effect of the MDPC($n$D/$m$L) and RS code respectively, no FEC is applied, neither to the main nor to the auxiliary channel. The BER curves for the individual channels are shown in Fig. \ref{mixfrequency}. As the BER performance of all 2.16 GHz channels is almost identical, the channel at a center frequency of 294.84 GHz has been selected as a representative. We observe that the larger the transmission distance, modulation order or frequency, the higher the BER. Apart from the BER, the data rate $D$ is computed based on the symbol rate and selected modulation scheme by
\begin{equation}
    D = \frac{\text{log}_2 M}{t_s} = \text{log}_2 M \cdot 2 \cdot B_N,
\end{equation}
where $M$ states the number of symbols of the modulation scheme, $t_s$ is the symbol duration and $B_N$ denotes the Nyquist-bandwidth. According to the IEEE802.15.3d standard, the Nyquist bandwidth of $B_{N,\text{2.16\,GHz}} = 880\,\text{MHz}$, $B_{N,\text{8.64\,GHz}} = 3520\,\text{MHz}$, $B_{N,\text{10.80\,GHz}} = 4400\,\text{MHz}$.

As the auxiliary and main channel use BPSK and 16QAM modulation at the same transmission distance, these modulation schemes have been selected for a comparison and plotted for the $5 \cdot 2.16$\,GHz and $ 1 \cdot 10.80$\,GHz reference systems mentioned above. The graphs of the reference systems for BPSK and 16QAM as well as the RS-code and MDPC protected systems are shown in Fig. \ref{BER_comparison}. As expected, it can be seen that the robust BPSK modulation performs much better than the less robust 16QAM modulation scheme. Both reference systems produce similar BER results for BPSK and 16QAM modulation schemes with little deviations. The RS protected system on the other hand performs better than the MDPC($n$D/$m$L) protected system in terms of their residual BER. Below $3.0$ m (MDPC-system) and below $3.5$ m (RS-system) transmission distance, the BER is not shown as it resulted in 0. The reference systems offer a BER of 0 only up to $2.0$ m transmission distance. There is a small improvement in the BER performance of the transmission due to the additional protection. This improvement for the main and auxiliary-channel system comes at the prize of a reduced throughput as part of the bandwidth is used for redundancy of the transmission. This can be seen in Fig. \ref{Goodput_comparison}. This figure shows the theoretical goodput, defined as the number of useful information bits delivered by the network to a certain destination per unit of time. The goodput is given by \begin{equation}
    G =R_F \sum_{i=1}^N D_{i} \cdot (1-BER_i) ,
\end{equation} whereby $N$ denotes the total number of channels of the THz system, $D_i$ is the data rate of channel $i$ and $BER_i$ represents BER of channel $i$. In general, the goodput decreases with an increasing transmission distance or increasing bandwidth, while it increases with an increasing modulation level. The curves show the same systems used for the evaluation in Fig. \ref{BER_comparison}. For error free transmission, the 16QAM systems can transfer 35 Gbps, while the BPSK systems offer only $8.75$ Gbps due to its lower spectral efficiency. The main and auxiliary channel system only uses $8.64$ GHz bandwidth for the transmission of information bits. In error-free conditions, it has only $80\%$ of the $10.80$ GHz system's throughput, hence 28 Gbps. As RS and MDPC($n$D/$m$L) offer an improved BER performance in comparison with the uncoded 16QAM transmissions due to their added redundancy, it can bee seen that the goodput declines less steep within the examined range up to 20\,m but remains below the uncoded system's values. Since BER of RS-system is lower than that of MDPC-system, the block error probability of RS-system is lower than that of MDPC-system as shown in Fig. \ref{error_block_for_coding}.

Next, in Fig. \ref{coderate_eve}, we theoretically analyze the code rate $R_F$ of two-channel THz system with coding vs. transmission distance $d_{main}$ of main channel, using MDPC($n$D/$m$L) or RS code. To get the code rate $R_F$ for a tolerant-fault THz system as analyzed above, we use Eq. \eqref{t_k}, Eq. \eqref{parameters-mn} and Eq. \eqref{coderate} for MDPC($n$D/$m$L) code; and apply Eq. \eqref{t_k_RS}, Eq. \eqref{p_s}, Eq. \eqref{parameters-RKforRS} and Eq. \eqref{coderate} for RS code, whereby the total number of error bits received at the receiver is equal to or as close as possible to the ability of correcting error of MDPC($n$D/$m$L) or RS code and the parameters of $K,R$ and $s$ chosen to satisfy these equations. We consider MDPC($n$D/$m$L) code with the ability of correcting error $t_{MDPC}=1$ bits per block, i.e., $n=2$, and RS code with the ability of correcting error $t_{RS}=1$ symbol per block. We note that if we choose a reasonable parameter $K$ as analyzed in section \ref{ana}, then it is unnecessary to increase the value of $t_{MDPC}$ or $t_{RS}$, which can cause decreasing the code rate as well as increasing coding and decoding complexity. The frequency of auxiliary channel chosen is as low as possible to be easier in setting its transmission distance so that its bit error probability $p_e^A$ is approximately equal to $0$, i.e., based on Fig. \ref{294dot84}, $d_{aux} \leq 8$ m, and the setting with $p_e^A=0$ is to increase the code rate for the system. The other parameters are similarly set as Fig. \ref{BER_comparison}. In general, the larger the transmission distance, the higher the error bit rate; the code rate $R_F$ decreases with an increasing transmission distance. On the other hand, the code rate of RS code is higher than that of MDPC($n$D/$m$L) code because increasing or decreasing the size of redundancy for MDPC($n$D/$m$L) code depends on the value of $K$ original bits, while the size of redundancy for RS code is fixed for all cases. In Fig. \ref{trans_overhead}, we theoretically analyze the transmission overhead $\theta$ of two-channel THz coding systems with the same configurations as Fig. \ref{coderate_eve}. We use Eq. \eqref{overhead} to calculate the transmission overhead $\theta$. In general, the larger the transmission distance, the higher the error bit rate; the transmission overhead $\theta$ increases with an increasing transmission distance. Additionally, the transmission overhead of RS code is lower than that of MDPC($n$D/$m$L) code. 

  \section{Conclusion}
In this paper we have proposed a two-channel THz system, including one THz main channel used to send original data and one THz auxiliary channel exploited to transfer parity data. Our analysis evaluated and compared this system with different configurations in term sof code rate, goodput, transmission overhead, expected residual BER and block error probability of MDPC($n$D/$m$L) and RS code required for a reliable system. The results showed that a small coding gain has to be achieved by trading in a portion of the available system bandwidth and hence reduces the overall transmission throughput. For the THz-transmission in the examined frequency ranges, the results suggest for AWGN that the system will have a positive effect by prolonging the error-free range for short range transmissions, however, the trade-off is significant and can likely be outperformed by lowering the code rate of a potential FEC scheme while keeping the throughput above the throughput of the main and auxiliary channel system. As a future work, the performance of the system should be evaluated under the consideration of forward error correction, realistic scenarios with multipath-propagation and hardware effects as all three are suspected to have an impact on the performances of the suggested system as well. 
\section*{Acknowledgment}
This work was partially supported by the DFG Project Nr. JU2757/12-1, "Meteracom: Metrology for parallel THz communication channels."
\bibliographystyle{IEEEtran}

\bibliography{nc-rest}

\end{document}